\documentclass[preprint, showpacs, aps, nofootinbib]{revtex4}

\usepackage{graphicx}
\usepackage{dcolumn}
\usepackage{bm}

\def\nslash{\rlap{\hspace{0.02cm}/}{n}}
\def\kslash{\rlap{\hspace{0.02cm}/}{k}}
\def\pslash{\rlap{\hspace{0.02cm}/}{p}}
\def\dslash{\rlap{\hspace{0.08cm}/}{D}}
\def\beq{\begin{eqnarray}}
\def\eeq{\end  {eqnarray}}

\def\pr{^{\prime}}

\def\non{\nonumber}

\def\as{\alpha_s}

\def\lqcd{\Lambda_{\rm QCD}}

\newcommand\npb{Nucl.\ Phys.\ B }

\newcommand\plb{Phys.\ Lett.\ B }

\begin{document}

\title{ The SCET$_{\rm II}$ and factorization }

\author{Zheng-Tao Wei \footnote{e-mail address: zwei@uv.es} }

\affiliation{ Departamento de F\'{\i}sica Te\'orica, Universidad
 de Valencia, \\ E-46100, Burjassot, Valencia, Spain }

\begin{abstract}
We reformulate the soft-collinear effective theory which includes
the collinear quark and soft gluons. The quark form factor is used
to prove that SCET$_{\rm II}$ reproduces the IR physics of the
full theory. We give a factorization proof in deep inelastic
lepton-hadron scattering by use of the position space formulation.
\end{abstract}

\pacs{12.38.Aw}

\maketitle

\newpage
\section{Introduction}

Perturbative QCD (pQCD) provides a powerful method to calculate
the hard processes with large momentum transfers $Q\gg \lqcd$ from
the first principle \cite{CSS, BL}. The key ingredient of pQCD is
factorization, which separates short-distance dynamics from the
long-distance physics. Recently, an effective filed theory for
soft and collinear particles were proposed in a series of studies
\cite{B1, B2, B3, B4, B5, B6}. The proof of the factorization
theorem in the soft-collinear effective theory (SCET) can be
performed at the operator level which is much simpler than the
diagrammatic analysis in pQCD. It also provides a new framework to
study the resummation of Sudakov double-logs and the power
corrections.

The SCET is divided into SCET$_{\rm I}$ and SCET$_{\rm II}$
according to the momentum of the collinear particles.  It is
convenient to use the light-cone coordinates $p^{\mu}=(p^+, p^-,
p_{\bot})$ where $p^+=n_-\cdot p, ~p^-=n_+\cdot p$ and $n_+, ~n_-$
are two light-like vectors which satisfy $n_+^2=n_-^2=0$ and
$n_+\cdot n_-=2$. For the process where the off-shellness of the
collinear particle $p_c^2\sim Q\lqcd$, the soft particle belongs
to ultrasoft mode. The momenta of the collinear and ultrasoft
particles are scaled as $p_c\sim Q(\lambda^2, 1, \lambda)$ and
$p_{us}\sim Q(\lambda^2, \lambda^2,\lambda^2)$ where
$\lambda=\sqrt{\lqcd/Q}$. The theory of describing the collinear
and ultrasoft fields is called SCET$_{\rm I}$. For the processes
where the off-shellness of the collinear particle $p_c^2\sim
\lqcd^2$, the soft particle belongs to soft mode. The momenta of
the collinear and soft particles are scaled as $p_c\sim
Q(\lambda^2, 1, \lambda)$ and $p_s\sim Q(\lambda,
\lambda,\lambda)$ where $\lambda=\lqcd/Q$. The SCET$_{\rm II}$
aims to describe the collinear and soft fields.

The construction of the SCET$_{\rm II}$ is considered more
complicated than the SCET$_{\rm I}$ because the momentum of the
collinear particle does not retain its scaling when a soft
particle couples to it. A view about the SCET$_{\rm II}$ is that
it can be considered as a low energy effective theory of
SCET$_{\rm I}$ and is obtained from the the SCET$_{\rm I}$ by
integrating out the ${\cal O}(\sqrt{Q\lqcd})$ fluctuations.
According to this view of point, in the leading order of
$\lambda$, there is no coupling of soft to collinear particles in
the SCET$_{\rm II}$. However, the above arguments of excluding the
interaction between the collinear and soft particles (in the
leading order) are insufficient. Recalling that in the Heavy-Quark
Effective Theory (HQET) \cite{Georgi}, although the off-shellness
of a heavy quark $p_Q^2=(m_Q v+k)^2\sim m_Q\lqcd$ is at a large
intermediate scale, the leading order HQET Lagrangian is given by
the interaction of the heavy quark and soft gluons as ${\cal
L}_{HQET}=\bar h_v iv\cdot D_s h_v$. The leading order interaction
of the collinear quark and soft gluon had been proposed in the
framework of the Large-Energy-Effective-Theory (LEET) \cite{DG}.
The LEET was criticized by lacking the collinear gluon degrees of
freedom in the development of the SCET. Because the interactions
of the collinear quark and soft gluon in the LEET is not
contradict to any physical principle, another way of constructing
the SCET$_{\rm II}$ can be done by adding the collinear gluon into
the LEET. This is our proposal. The construction of the
interactions of the collinear quark and soft gluons in the leading
order at the Lagrangian level is more transparent and it is easier
to be extended to higher orders.

We will study the quark form factor in the asymptotic limit in the
SCET$_{\rm II}$. First, we will show that the SCET$_{\rm II}$
reproduce the IR physics of QCD for the  quark form factor at
one-loop order. Then we use the SCET$_{\rm II}$ to give a
factorized form for the quark form factor in the asymptotic limit.
Another motivation of this study is to explore the proof of
factorization by using the position space representation of SCET.
In the position space representation, one can use the conventional
definition of the universal non-perturbative quantities. As an
example, we discuss the factorization proof in deep inelastic
lepton-hadron scattering (DIS) process.

\section{The SCET$_{\rm II}$ Lagrangian}
\label{sec:SCET}

It is convenient to use the light-cone coordinates to study the
processes which contain highly energetic light hadrons or jets. An
arbitrary four-vector $p^{\mu}$ is written as %
\beq %
p^{\mu}=(p^+, p^-, p_{\bot})=n_-\cdot p\frac{n_+^{\mu}}{2}+
        n_+\cdot p\frac{n_-^{\mu}}{2}+p_{\bot}^{\mu},
\eeq %
where $n_+^{\mu}=(2,0,0_{\bot})$ and $n_-^{\mu}=(0,2,0_{\bot})$
are two light-like vectors which satisfy $n_+^2=n_-^2=0$, and
$n_+\cdot n_-=2$. We discuss a case that all collinear particles
move close to $n_-$ direction, i.e., the momentum $p^-\sim Q$ is
large at first for the simplification of illustration. Other cases
can be similarly obtained. A four-component Dirac field $\psi$ can
be decomposed into two-component spinors $\xi$ and $\eta$ by %
\beq %
\psi=\xi+\eta, ~~~~ %
\xi \equiv P_-\psi=\frac{\nslash_-\nslash_+}{4}\psi, ~~~~%
\eta\equiv P_+\psi=\frac{\nslash_+\nslash_-}{4}\psi.
\eeq %
with $\nslash_-\xi=\nslash_+\eta=0$. We will consider the
effective theory for massless quarks because the mass of the light
$u,d $ quarks are much smaller than the QCD confinement scale
$\lqcd$. The massless QCD Lagrangian
${\cal L}=\bar{\psi}~i\dslash \psi$ is%
\beq %
\label{eq: QCD1} %
{\cal L}=\bar{\xi }~in_-\cdot D\frac{\nslash_+}{2}~\xi +
         \bar{\eta}~in_+\cdot D\frac{\nslash_-}{2}~\eta+
         \bar{\xi}~i\dslash_{\bot}\eta+
         \bar{\eta}~i\dslash_{\bot}\xi.
\eeq %
where
$D_{\mu}=\partial_{\mu}-igA_{\mu}=\partial_{\mu}-igT^aA_{\mu}^a$.

As we have discussed in the Introduction, the SCET$_{\rm II}$
describes the degrees of freedom of the collinear and soft fields.
The momenta for the collinear and soft particles are scaled as
$p_c\sim Q(\lambda^2, 1, \lambda)$ and $p_s\sim Q(\lambda,
\lambda,\lambda)$ where $\lambda=\lqcd/Q$. The power counting for
the relevant fields can be found in \cite{BCDF}: the collinear
quark field $\xi\sim \lambda$; the collinear gluon field:
$A_c^{\mu}\sim (\lambda^2,1,\lambda)$; the soft quark field:
$q_s\sim \lambda^{3/2}$; the soft gluon field: $A_s\sim \lambda$.

Now, we follow the construction of the LEET given in \cite{DG,
CYOPR} and then transform it into the position space
representation. After the interactions with the soft gluon, the
momentum of the collinear quark is changed into $p=p_c+k$ where
$k\sim Q(\lambda, \lambda, \lambda)$ is the residual momentum
carried by the soft gluons. Thus, the scaling of the integral
element is $d^4 x\sim \frac{1}{d^4 p}\sim \lambda^{-3}$ rather
than $\lambda^{-4}$. Analogous to the HQET, we remove the large
momentum component $p^-$ by redefining the collinear quark field
as $\psi(x)=e^{-i\frac{(n_+\cdot p)}{2} n_-\cdot
x}[\xi_{p^-}(x)+\eta_{p^-}(x)]$. After integrating out the
small-component field $\eta_{p^-}(x)$, one obtain the effective
Lagrangian for the interaction of collinear
quark with the soft gluons as%
\beq \label{eq:cs} %
{\cal L}_{c-s}=\sum_{p^-}\bar{\xi}_{p^-}\left [~in_-\cdot D_s
  +i\dslash_{s\bot}\frac{1}{~n_+\cdot p+in_+\cdot D_s}~ i\dslash_{s\bot}
  ~\right ] \frac{\nslash_+}{2}~\xi_{p^-}.
\eeq %
where $D_s=\partial_s-igA_s$ and $\partial_s$ scales as a soft
momentum.  The lowest order effective Lagrangian for the collinear
quark and soft gluon interaction is %
\beq %
\label{eq:QCDc-s} {\cal L}_{c-s}^{(0)}=\sum_{p^-}\bar{\xi} _{p^-}
   ~in_-\cdot D_s\frac{\nslash_+}{2}~\xi_{p^-}.
\eeq %
Eq. (\ref{eq:QCDc-s}) is the LEET Lagrange proposed in \cite{DG}.
Eq. (\ref{eq:cs}) also provides the higher order terms which is
given in the second term of it. Since only the $n_-\cdot A_s$
gluons contribute at the lowest order, one can write
$\xi_{p^-}(x)$ field in another way as%
\beq \label{eq:Ws}%
\xi_{p^-}(x)=W_s(x)~\xi^0_{p^-}(x), ~~~~~ W_s(x)={\rm P~exp}\left
( ig\int_{-\infty}^0 dt ~n_-\cdot A_s(x+tn_-) \right ).
\eeq %
The field $\xi^0$ is a quark field without coupling to the soft
gluons. The ${\rm P}$ in the soft Wilson line $W_s$ denotes
path-ordering defined such that the the gluon fields stand to the
left for larger values of parameter $t$.

The previous obstruction of constructing the leading order
interaction of the collinear quark with soft gluons in the
SCET$_{\rm II}$ lies in the different scaling of momenta $p_c$ and
$p_c+k_s$. In the above analysis, the main assumption we use is
that $p^-$ is the largest momentum component. The $\xi_{p^-}(x)$
field is a collinear field rather than a soft field like the heavy
quark field $h_v(x)$. This can be seen from Eq. (\ref{eq:Ws}). The
off-shellness of $p_c+k_s$ does not invalidate our assumption. It
is the superselection rule of momentum component $p^-$ (the energy
of the collinear particle) that permits us to write down the
lowest order effective interaction of Eq. (\ref{eq:QCDc-s}) at the
Lagrangian level. Other interaction, such as the collinear gluon
coupling to the heavy quark can not can not been done in this way.
From another point of view, the information of the off-shellness
cause by $p^-k^+$ can be included in the soft Wilson line, so the
construction of the leading order effective Lagrange is equivalent
to the use of the soft Wilson line. The advantage of our proposal
is that it permits us to write down not only the leading order
term but also the higher order terms systematically.

Transforming the above analysis into the position space
formulation can be done straightforwardly by using the
correspondence rule $p\to i\partial$. The index of $p^-$ in the
field $\xi_{p^-}$ can be dropped. We also include the collinear
gluon in the effective theory. The procedure is done by
integrating out the $\eta$ field and then insert the solution into
Eq. (\ref{eq: QCD1}) as in \cite{BCDF}, we obtain an effective
Lagrangian for SECT$_{\rm II}$ as%
\beq \label{eq:SCET}%
{\cal L}_{\rm SCET_{II}}=\bar{\xi}\left [ ~in_-\cdot D
  +i\dslash_{\bot}\frac{1}{~in_+\cdot D}~ i\dslash_{\bot}
  \right ]\frac{\nslash_+}{2}~\xi.
\eeq %
where the covariant derivative is defined by
$D=\partial-igA_c-igA_s$. The Eq. (\ref{eq:SCET}) contains the
interaction of the collinear quark with  collinear gluon %
\beq %
{\cal L}_{c-c}=\bar{\xi}\left [ ~in_-\cdot D_c
  +i\dslash_{c\bot}\frac{1}{~in_+\cdot D_c}~ i\dslash_{c\bot}
  \right ]\frac{\nslash_+}{2}~\xi.
\eeq %
The interaction of the collinear quark with soft gluon is given in
Eq. (\ref{eq:cs}). The new interaction terms of the collinear
quark, collinear gluon and soft gluon appear, such as %
${\cal L}_{s-c}=\bar{\xi}\left [
  i\dslash_{s\bot}\frac{1}{~in_+\cdot D_c}~ i\dslash_{c\bot}
 +i\dslash_{c\bot}\frac{1}{~in_+\cdot D_c}~ i\dslash_{s\bot}
  \right ]\frac{\nslash_+}{2}~\xi$ which are $\lambda$ order
corrections. One must keep in mind that the scaling of the
integral element $d^4 x$ is $\lambda^4$ for the
collinear-collinear interaction and $\lambda^3$ for collinear-soft
interaction. The full SCET$_{\rm II}$ should contain the soft
quarks as well. Some discussions about this part can be found in
\cite{HN}. We will not consider it in this study.

One can define the collinear Wilson line $W_c$  in a similar way
as the soft Wilson line by %
\beq \label{eq:Wc2}%
W_c(x)={\rm P~exp}\left ( ig\int_{-\infty}^0 dt ~n_+\cdot
       A_c(x+tn_+) \right ).
\eeq %
We will use the collinear Wilson line in the proof of
factorization below.

The SCET contains rich symmetry structures. Except the symmetries
which had been studied largely in the literatures, we discuss a
new symmetry: scale symmetry. However, this symmetry is not
rigorous but approximate. It is broken by the renormalization
effect and quark mass. In the SCET$_{\rm II}$, the light quark
mass is approximated to zero.\footnote{For s quark, the mass
effect is not negligible.} In most high energy hard processes, the
broken of the scale symmetry is caused by the perturbative
corrections occurred at the scale of ${\cal O}(Q)$ which is
suppressed by the coupling constant $\alpha_s(Q)$. If neglecting
the perturbative corrections, the scale symmetry can be a useful
guide. For the SCET, the effective Lagrangian at tree level is
scale invariant. The scale
transformation is defined as%
\beq %
x_{\mu}\to s^{-1} x_{\mu}, ~~~~~
\partial_{\mu}\to s~\partial_{\mu}, ~~~~~
\xi\to s^{3/2}\xi,   ~~~~~
A_{\mu}\to s A_{\mu}.
\eeq %
It is easy to check that the action $\int d^4x {\cal L}_{\rm
SCET_{II}}$ is invariant under the above scale transformation.

\section{The quark form factor}

The quark form factor in the asymptotic limit provides a simple
example to discuss the Sudakov resummation and factorization
\cite{Collins}. We use this example to check whether SCET$_{\rm
II}$ reproduces all the IR physics of QCD. We consider a form
factor given by $\langle B|\bar\psi_B\gamma_{\mu}\psi_A |A
\rangle=\bar u_B\gamma_{\mu}u_A F$ in the full QCD. The current is
electromagnetic. The process is that an initial energetic quark A
absorbs a highly off-shell photon and transforms into a final
energetic quark B in the opposite direction. We choose the
transition form factor rather than the annihilation form factor in
order to avoid the unimportant imaginary part. We study the case
that the collinear quarks are both on-shell.  The momenta of
quarks A and B are chosen as $p_A=(Q, 0, 0_{\bot}), ~p_B=(0, Q,
0_{\bot})$ and $q^2=(p_B-p_A)^2=-Q^2$ where $Q$ is a large energy
scale.

The one-loop order QCD corrections to the quark transition form
factor contain the quark self-energy and the vertex corrections.
The light quark self-energy corrections are same in the full
theory and the effective theory \cite{B2}. We will not consider
them below. The one-loop vertex correction depicted in Fig. 1
contains the ultraviolet (UV) divergences\footnote{It is cancelled
by the quark field renormalization.} as well as the collinear and
soft divergences. The collinear divergence appears when the
virtual gluon becomes collinear to quark A or B. One may use the
quark mass and a fictitious mass for gluon field to regulate the
IR singularities. The problem of this regularization method is
that it is insufficient to regulate all the singularities in the
effective theory. We will use the regularization method proposed
in \cite{BDS} because it simplifies the calculation. The IR
regulator is given by adding new terms in the Lagrangian: %
\beq %
{\cal L}_{reg}=
  -\frac{\delta}{2}A_{+\mu}(in_-\cdot \partial) A^{\mu}_+
  -\frac{\delta}{2}A_{s\mu}(in_-\cdot \partial) A^{\mu}_s
  +  (+ \leftrightarrow -).
\eeq %
where $A_+$ represents the gluon collinear to the A quark. The
appearance of the additional terms $(+ \leftrightarrow -)$ is
because there are two collinear quarks with different directions
in our process.

\begin{figure}
\includegraphics[scale=1.0]{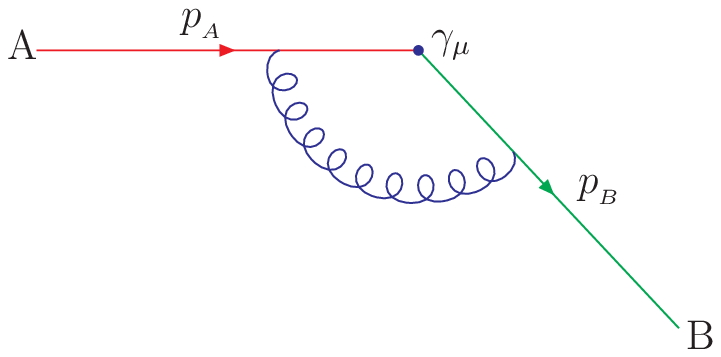}
\caption{The one-loop vertex correction to the quark form factor.}
\end{figure}

We use the dimensional regularization to regulate the UV
divergences and perform the calculation in $d=4-2\epsilon$
dimension. The one-loop vertex correction in the full theory
plotted in Fig. 1 is given by %
\beq \label{eq:fulli}%
I_{full}=-ig_s^2 C_F\mu^{\prime 2\epsilon}\int\frac{d^d k}
  {(2\pi)^d} \frac {\gamma_{\rho}(\kslash+\pslash_B)
  \gamma_{\mu}(\kslash+\pslash_A) \gamma^{\rho}}
  {[(k+p_A)^2+i\epsilon][(k+p_B)^2+i\epsilon]
  [k^2-\delta(n_++n_-)\cdot k+i\epsilon]}.
\eeq %
The calculation of the above integral can be done in the
conventional way and the full theory result is %
\beq %
I_{full}=\frac{\as}{4\pi}C_F \gamma_{\mu}\left[ \frac{1}{\epsilon}
 +{\rm ln}\frac{\mu^2}{Q^2}-\frac{1}{2}{\rm ln}^2\frac{Q^2}{\delta^2}
 +2{\rm ln}\frac{Q^2}{\delta^2}-\frac{\pi^2}{3} \right]
\eeq %
where $\mu^2=4\pi \mu^{\prime 2}{\rm e}^{-\gamma_E \epsilon}$ and
$\gamma_E$ is the Euler constant.

Maybe the best way to check that SCET reproduces the IR physics of
the full QCD is performed by using the method of regions
\cite{BS}. The idea is expanding the integrand $I$ in the momentum
regions which give contributions in dimensional regularization. In
the quark form factor, the non-vanishing contributions come from
the hard region where the momentum of the virtual gluon $k\sim
(1,1,1)$,  collinear-A region where $k\sim (1, \lambda^2,
\lambda)$, collinear-B region where $k\sim (\lambda^2, 1,
\lambda)$ and  soft region where $k\sim (\lambda, \lambda,
\lambda)$.

For the hard region, we can set $\delta=0$. The hard region
contribution is%
\beq %
I_h&=&-ig_s^2 C_F\mu^{\prime 2\epsilon}\int\frac{d^d k}
  {(2\pi)^d} \frac {\gamma_{\rho}(\kslash+\pslash_B)
  \gamma_{\mu}(\kslash+\pslash_A) \gamma^{\rho}}
  {[k^2+2p_A\cdot k+i\epsilon][k^2+2p_B\cdot k+i\epsilon]
  [k^2+i\epsilon]} \non \\
  &=&\frac{\as}{4\pi}C_F \gamma_{\mu}\left[ -\frac{2}{\epsilon^2}
  -\frac{3+2{\rm ln}\frac{\mu^2}{Q^2}}{\epsilon}
 -{\rm ln}^2\frac{\mu^2}{Q^2}-3{\rm ln}\frac{\mu^2}{Q^2}
 -8+\frac{\pi^2}{6} \right]
\eeq %

The collinear-A region comes from the contribution where the
momentum $k$ is parallel to $p_A$. We make a transformation $k\to
-k$ in Eq. (\ref{eq:fulli}). In this region,
$\frac{\gamma_\rho(\pslash_B-\kslash)}{k^2-2p_B\cdot
k+i\epsilon}\to \frac{-n_{-\rho}}{n_-\cdot k}$. The collinear-A
region contribution is%
\beq %
I_A&=&-ig_s^2 C_F\mu^{\prime 2\epsilon}\int\frac{d^d k}
  {(2\pi)^d} \frac {-n_{-\rho}
  \gamma_{\mu}(\pslash_A-\kslash) \gamma^{\rho}}
  {[k^2-2p_A\cdot k+i\epsilon]~(n_-\cdot k)~
  [k^2-\delta (n_-\cdot k)+i\epsilon]} \non \\
  &=&\frac{\as}{4\pi}C_F \gamma_{\mu}\left[ \frac{2}{\epsilon^2}
  +\frac{2+2{\rm ln}\frac{\mu^2}{Q \delta}}{\epsilon}
 +{\rm ln}^2\frac{\mu^2}{Q\delta}+2{\rm ln}\frac{\mu^2}{Q\delta}
 +4-\frac{\pi^2}{6} \right]
\eeq %
The above calculation is performed by the contour integration.

The collinear-B region contribution can be obtained from $I_A$ by
$ -\to +$ and $A\to B$. Since the final result of $I_A$ does not
depend on the changes, the collinear-B region contribution is the
same as $I_A$. Thus,%
\beq %
I_B&=&-ig_s^2 C_F\mu^{\prime 2\epsilon}\int\frac{d^d k}{(2\pi)^d}
  \frac{ \gamma_{\rho}(\pslash_B-\kslash)\gamma_{\mu} (-n_+^{\rho})}
  {[k^2-2p_B\cdot k+i\epsilon]~(n_+\cdot k)~
  [k^2-\delta (n_+\cdot k)+i\epsilon]} \non \\
  &=&\frac{\as}{4\pi}C_F \gamma_{\mu}\left[ \frac{2}{\epsilon^2}
  +\frac{2+2{\rm ln}\frac{\mu^2}{Q \delta}}{\epsilon}
 +{\rm ln}^2\frac{\mu^2}{Q\delta}+2{\rm ln}\frac{\mu^2}{Q\delta}
 +4-\frac{\pi^2}{6} \right]
\eeq %

The soft region contribution is %
\beq %
I_s&=&-ig_s^2 C_F\mu^{\prime 2\epsilon}\int\frac{d^d k}{(2\pi)^d}
  \frac{ n_{-\rho}\gamma_{\mu} n_+^{\rho}}
  {[n_-\cdot k+i\epsilon][n_+\cdot k+i\epsilon]
  [k^2-\delta (n_++n_-)\cdot k+i\epsilon]} \non \\
  &=&\frac{\as}{4\pi}C_F \gamma_{\mu}\left[ -\frac{2}{\epsilon^2}
  -\frac{2{\rm ln}\frac{\mu^2}{\delta^2}}{\epsilon}
  -{\rm ln}^2\frac{\mu^2}{\delta^2}-\frac{\pi^2}{6} \right]
\eeq %
The detailed calculation of the above results and the summation of
the Sudakov double-logs  will be given elsewhere.

The contributions from the collinear-A, collinear-B and soft
regions are infrared physics. They also constitute the results in
the SECT$_{\rm II}$ exactly.  It can be checked that
$I_{full}=I_h+I_A+I_B+I_s$. Thus, the IR physics of the quark form
factor is reproduced in the SCET$_{\rm II}$ and the non-IR physics
(it refers to the perturbative part in pQCD) is contained in the
hard region contribution.

After the proof that SCET$_{\rm II}$ reproduces the IR physics of
the quark form factor in the full theory, we can use the effective
theory to give a factorization form. The advantage of using the
SCET to study factorization is that we can use the operator
language rather than diagram analysis and the proof to all orders
can be done simply. The light-light current in the full QCD is:
$\bar\psi_B \gamma_{\mu}\psi_A$; in the
SCET$_{\rm II}$, the current operator becomes as%
\beq %
\bar\xi_{n_-}W_c(n_-)\gamma_{\mu}W_c^{\dagger}(n_+)\xi_{n_+}
   W_s^{\dagger}(n_-)W_s(n_+)
\eeq %
where $\xi_{n_-}$ and $W_c(n_-)$ represents the collinear field
and collinear Wilson line along the $n_-$ direction. Similar
definitions for the other operators are implied.

The factorized form for the quark form factor is%
\beq %
F=H\times J_A \times J_B\times S
\eeq %
where %
$J_A\equiv \langle 0|W_c^{\dagger}(n_+)\xi_{n_+}|A\rangle$,
$J_B\equiv \langle B|\bar\xi_{n_-}W_c(n_-)      |0\rangle$ and %
$S  \equiv \langle 0|W_s^{\dagger}(n_-)W_s(n_+) |0\rangle$. Our
definition of the jet-like and soft functions are different from
that of \cite{Collins}. The author in \cite{Collins} may adopt a
momentum subtraction to suppress the non-analytic terms in the
integral of a momentum region in order to avoid the double
counting. In the method of regions, these non-analytic terms
cancel in the sum of the integrals for different momentum regions.

\section{The factorziation in DIS}

The factorization is the foundation of pQCD but to prove it is
usually difficult. The deep inelastic lepton-hadron scattering
(DIS) plays a central role in understanding the factorization in
pQCD. In \cite{CSS}, the diagrammatic analysis is used to prove
the factorization to all orders of coupling constant $\alpha_s$
and leading order in $\lambda$ (or say $\lqcd/Q$). The method of
regions in \cite{BS} may be also viewed as the diagrammatic
analysis but a refined version which permits the analysis to
higher orders in $\lambda$. As we have discussed in the quark form
factor, the contributions of different regions (except for the
hard region) are corresponding parts in the SCET. After we have
checked that SCET produces all the IR physics of the full theory,
we can use the operator language in SCET in place of the
diagrammatic analysis to simplify the all-orders proof of
factorization.

The study of the factorization in DIS in the framework of SCET has
been given in \cite{B6}. 
There is a view that the effective theory used is SCET$_{\rm I}$
for the inclusive processes and SCET$_{\rm II}$ for the exclusive
processes. In \cite{B6}, the authors apply the SCET$_{\rm I}$ to
study the inclusive DIS process. This view is not rigorous. Which
SECT theory is used as the final effective theory of a process
depends on the momentum of collinear particle. In DIS, the proton
and the collinear quark inside the proton have the virtuality of
order of $\lqcd$, so the correct theory needed is SCET$_{\rm II}$
rather than SCET$_{\rm I}$. The analysis of \cite{B6} used a
hybrid position-momentum space representation in which a
definition of the parton distribution functions in the momentum
space is introduced. We will use the position space representation
developed in \cite{BCDF} and the SCET$_{\rm II}$ to study the
factorization in DIS. We can use the familiar definition of the
parton distribution functions in the position space. We think our
formalisms are more transparent and simpler than that in the
hybrid position-momentum representation.

We consider the DIS process $e(k)+H(p)\to e(k\pr)+X$. The
kinematical variable are defined as $q^2=-Q^2$ and
$x=\frac{Q^2}{2p\cdot q}$ with $q=k\pr-k$ and $p, m_H$ are the
momentum and mass of the initial hadron. We are interested in the
region that $Q\gg \lqcd$ and $x$ is fixed but not small. We choose
the frame so that $p^{\mu}=(0, \frac{Q}{x}, 0_{\bot})$ and
$q^{\mu}=(Q, 0, q_{\bot})$ with $q_{\bot}^2=Q^2$. Note that our
choice of frame is not same as that in \cite{B6}.

The hadronic tensor of the DIS cross section is determined by %
\beq \label{eq:wmunu1}%
W_{\mu\nu}=\left(-g_{\mu\nu}+\frac{q_{\mu}q_{\nu}}{q^2}\right)
 F_1(x, Q^2)+
 \frac{(p_{\mu}+\frac{q_{\mu}}{2x})(p_{\nu}+\frac{q_{\nu}}{2x})}
 {p\cdot q}F_2(x, Q^2)
\eeq %
The $F_{1,2}$ are the standard structure functions. The hadronic
tensor $W_{\mu\nu}$ is related to the forward scattering amplitude by%
\beq \label{eq:wmunu2}%
W_{\mu\nu}=\frac{1}{2\pi}\frac{1}{2}\sum_{\rm spin}\int d^4
  y~e^{iq\cdot y} \langle H|T[j_{\mu}(y)j_{\nu}(0)]|H \rangle
\eeq %
where $j_{\mu}$ is the electromagnetic current. In the discussions
below, we will not write the spin average $\frac{1}{2}\sum_{\rm
spin}$ for simplification. One can obtain the structure functions
through the projecting of hadronic tensor by
\beq \label{eq:projection}%
n_+^{\mu}n_+^{\nu}W_{\mu\nu}=\frac{2}{x}F_2, ~~~~~~
n_-^{\mu}n_-^{\nu}W_{\mu\nu}=\frac{1}{2x}F_2-F_1.
\eeq %

In the quark form factor, the leading contributions come from the
hard, collinear and soft regions. For DIS, the soft divergences
are cancelled due to the unitarity of Wilson line
$W_s^{\dagger}(n_-)W_s(n_-)$. The collinear quarks are moving in
the same direction in DIS which is different from the case in the
quark form factor. So, there is no soft factor appeared in the
factorization form in leading order. Physically, the cancellation
of soft divergences is the result of summing over all final
states.

Now, we consider the operator product $T[j_{\mu}(y)j_{\nu}(0)]$.
This product is light-cone dominated, i.e., $0\leq y^2 \leq
\frac{\rm const.}{Q^2}$. We can not apply the conventional short
distance operator-product-expansion \cite{Wilson} to disentangle
short-distance from long-distance dynamics. The light cone
expansion was proposed to study the light-cone dominated
processes. In SCET, one can use the SCET operators to expand the
the operator product. We have shown that the SCET$_{\rm II}$
reproduces the IR physics of the full theory in the quark form
factor. We expect that this conclusion can be applied in DIS.
Thus, we can expand the operator product $T[j_{\mu}(y)j_{\nu}(0)]$
in terms of the SCET operators. Take
$n_+^{\mu}n_+^{\nu}T[j_{\mu}(y)j_{\nu}(0)]$ as an
example of illustration %
\beq \label{eq:OPE}%
n_+^{\mu}n_+^{\nu}T[j_{\mu}(y)j_{\nu}(0)] \longrightarrow \sum_i
  C_i(y)O_i(yn_+).
\eeq %
where $O_i$ are operators in SCET and $yn_+\equiv (n_-\cdot
y)\frac{n_+}{2}$. In leading order of $\lambda$,
the gauge invariant SCET operators which contain collinear quark
and collinear gluon for DIS are given by%
\beq \label{eq:operator}%
O_i^q=(\bar\xi W_c)(yn_+)\frac{\nslash_+}{2}W_c^{\dagger}\xi(0),
~~~~~~ %
O_i^g=\sum_{j=1}^{2}(F^{+ j}W_c)(yn_+)W_c^{\dagger}F^+_{~j}(0).
\eeq %
where $F^{ij}$ is the gluon field strength operator. The
appearance of the gluon operator starts from $\alpha_s$ order. The
appearance of bilocal operators is the summation of infinite local
operators of leading order (order $\lambda^2$). The physical
interpretation is that the intermediate quark in the operator
product can emit (or absorbs) a large number of collinear gluons
in leading order. The Wilson coefficients $C_i$ in Eq.
(\ref{eq:OPE}) are infrared finite because it is obtained from the
matching from the full QCD onto the SCET$_{\rm II}$. The Wilson
coefficients contain the contribution from the hard region only
and is insensitive to the IR physics in principle.

According to \cite{CSS}, the hadronic matrix elements of the
bilocal operator are defined by %
\beq \label{eq:PDF}%
\langle H(p)|\bar\xi (yn_+) \frac{\nslash_+}{2}{\cal P}
  \xi(0)|H(p)\rangle&=&2(n_+\cdot p)
  \int d\zeta ~e^{i\zeta p^-y^+}f_{q/H}(\zeta),\non \\
\langle H(p)|F^{+ j}(yn_+){\cal P}
  F^+_{~j}(0)|H(p)\rangle&=&\zeta(n_+\cdot p)^2
  \int d\zeta ~e^{i\zeta p^-y^+}f_{g/H}(\zeta).
\eeq %
where ${\cal P}=W_c(yn_+)W_c^{\dagger}(0)$ is the ordered
exponential. $f_{q(g)/H}$ are parton distribution function for
quarks or gluons. The variable $\zeta$ is the light-cone momentum
fraction.

Introduce short-distance Wilsion coefficients in momentum space
$\tilde{C}_i(\zeta, x, Q)$ through the Fourier transformation by%
\beq \label{eq:coeff}%
\tilde{C}_i(\zeta, x, Q)=\int d^4y~e^{i\left( q+\zeta (p\cdot
  n_+)n_- \right )\cdot y}C_i(y)
\eeq %
Inserting Eqs. (\ref{eq:OPE}, \ref{eq:operator}, \ref{eq:PDF},
\ref{eq:coeff}) back into Eqs. (\ref{eq:wmunu1}, \ref{eq:wmunu2},
\ref{eq:projection}), we obtain a convolution factorization
formula for the structure functions %
\beq %
F_{1,2}(x, Q^2)=\sum_a\int_x^1 d\zeta~f_{a/H}(\zeta,\mu)
  H_{1,2}^a(\zeta, x, Q/\mu)
\eeq %
where $a$ represents quark or gluon and the hard scattering kernel
$H_{1,2}^a(\zeta, x, Q/\mu)$ differs from $\tilde{C}_i(\zeta, x,
Q/\mu)$ by constant factors. The factorization formula is a
convolution form because both the parton distribution functions
and the Wilson coefficients depend on $\zeta$. Thus, we proved
that the structure functions can be factorized into a convolution
of short-distant and long-distant parts.

\section{Discussions and conclusions}

The study of the hard QCD processes has been a long history. The
development of soft-collinear effective theory may make these
processes more understandable. In this study, we reformulated the
SCET$_{\rm II}$ along the line of the LEET. We give the leading
order interaction of the collinear quark with soft gluons at the
Lagrangian level. The higher order terms can be obtained from the
effective Lagrangian. In our approach, the differences between
SCET$_{\rm II}$ and the SCET$_{\rm I}$ are less than the previous
studies. According to this view, most of the results in SCET$_{\rm
I}$ may be applied into the SCET$_{\rm II}$. But one keep in mind
that the power counting of the soft-collinear interaction in
SCET$_{\rm II}$ are different from the ultrasoft-collinear part in
SCET$_{\rm I}$. The more detailed study about this is necessary.

The quark form factor is used as an example to prove that
SCET$_{\rm II}$ does reproduce all the IR physics of the full QCD.
We studied the factorization of the quark form factor and DIS in
SCET$_{\rm II}$. Obvious, our proof is not full, but it contains
the main ingredients. Although which representation is more
convenient is not easy to evaluate, the position space
representation is more simple and intuitive in the proof of
factorization in our opinion. The SCET also provide a new
framework to study the processes where the conventional
operator-product-expansion is not applicable.

Although SCET has been studied in some literatures, we feel that
the construction of SCET$_{\rm II}$ and its application into the
processes where soft gluon does not cancel is still a difficult
task. The study of exclusive B decays and other high energy
processes such as diffractive scattering require better
understanding of the SCET$_{\rm II}$.

\section*{Acknowledgments}

It is a pleasure to thank J. Bernabeu, J. Papavassiliou, F.
Campanario and M. Nebot for many useful discussions. I also thank
S. Fleming for discussions on conference FPCP2003. The author
acknowledges a fellowship of the Spanish Ministry of Education.
This research has been supported by Grant FPA/2002-0612 of the
Ministry of Science and Technology.


\end{document}